\documentclass[superscriptaddress,showkeys,twocolumn,nofootinbib,longbibliography]{revtex4-1}

\usepackage{amsmath}
\usepackage{amssymb}
\usepackage{graphicx}
\usepackage{epsf}
\usepackage{slashed}
\usepackage{enumitem}
\usepackage[czech,english]{babel}
\usepackage[cp1250]{inputenc}
\usepackage{hyperref}
\usepackage{xcolor}

\usepackage[only,llbracket,rrbracket,llparenthesis,rrparenthesis]{stmaryrd} % for comparison with 4 similar parenthesis symbols
\usepackage{accsupp} % for ensuring the right Unicode codepoint upon pasting

\newcommand{\I}{\ensuremath{\mathrm{i}}}% imaginary unit
\newcommand{\e}{\ensuremath{\mathrm{e}}}% Euler number
\newcommand{\cc}{\ensuremath{\mathrm{c.c.}}}% complex conjugated part
\newcommand{\abs}[1]{\left|#1\right|}

\newcommand{\refer}[1]{(\ref{#1})}

\usepackage{bbm} %\mathbbm{1}, \mathbbmss{1}, \mathbbmtt{1}

\usepackage[nodayofweek]{datetime}
\newdateformat{mydate}{\twodigit{\THEDAY}{ }\shortmonthname[\THEMONTH], \THEYEAR}
\usepackage[normalem]{ulem}
\usepackage{color}
\usepackage{ulem} % podtrhavani, skrtani atd. textu \uline,\uuline,\uwave,\sout,\xout

\definecolor{myred}{rgb}{1,0,0}
\definecolor{mygreen}{rgb}{0,0.8,0.2}
\definecolor{myblue}{rgb}{0,0,1}

\definecolor{Ared}{rgb}{1,0.7,0}
\definecolor{Agreen}{rgb}{0.7,0.8,0.2}
\definecolor{Ablue}{rgb}{0,0.7,1}

\renewcommand{\emph}[1]{\textit{#1}}

\begin{document}

\definecolor{mygreen}{HTML}{006E28}
\newcommand{\kasia}[1]{{\color{mygreen}\textbf{?KS:}  #1}}
\newcommand{\trom}[1]{{\color{red}\textbf{?TR:} \color{red} #1}}
\newcommand{\filip}[1]{\textbf{?FB:} {\color{blue} #1}}
\newcommand{\aw}[1]{{\color{magenta}\textbf{?AW:}  #1}}
%\preprint{APS/123-QED}

\title{Oscillons from $Q$-balls through renormalization}
\author{F. Blaschke}

\affiliation{Research Center for Theoretical Physics and Astrophysics, Institute of Physics, Silesian University in Opava, Bezru\v{c}ovo n\'{a}m\v{e}st\'{\i}~1150/13, 746~01 Opava, Czech Republic
}
\affiliation{Institute of Experimental and Applied Physics, Czech Technical University in Prague, Husova~240/5, 110~00 Prague~1, Czech Republic}

\author{T. Roma\'{n}czukiewicz}
%\email[]{tomasz.romanczukiewicz@uj.edu.pl}
\affiliation{
 Institute of Theoretical Physics,  Jagiellonian University, Lojasiewicza 11, 30-348 Krak\'{o}w, Poland
}

\author{K. S\l{}awi\'{n}ska}
%\email[]{tomasz.romanczukiewicz@uj.edu.pl}
\affiliation{
 Institute of Theoretical Physics,  Jagiellonian University, Lojasiewicza 11, 30-348 Krak\'{o}w, Poland
}

\author{A. Wereszczy\'{n}ski}
%\email[]{andrzej.wereszczynski@uj.edu.pl}

\affiliation{
 Institute of Theoretical Physics,  Jagiellonian University, Lojasiewicza 11, 30-348 Krak\'{o}w, Poland
}
 \affiliation{Department of Applied Mathematics, University of Salamanca, Casas del Parque 2, 37008 - Salamanca, Spain
}

\affiliation{International Institute for Sustainability with Knotted Chiral Meta Matter (WPI-SKCM2), Hiroshima University, Higashi-Hiroshima, Hiroshima 739-8526, Japan}

\begin{abstract}
Using  a renormalization-inspired perturbation expansion we show that oscillons in a generic field theory in (1+1) dimensions arise as dressed $Q$-balls of a universal (up to the leading nonlinear order) complex field theory. This theory reveals a close similarity to the integrable complex sine-Gordon model which possesses exact multi-$Q$-balls. We show that excited oscillons, with characteristic modulations of their amplitude, are two-oscillons bound states generated from a two $Q$-ball solution.
\end{abstract}

\maketitle

%%%%%%%%%%%%%%%%%%%%%%%%%%%%%%%%%%%%%
\section{Introduction}
%%%%%%%%%%%%%%%%%%%%%%%%%%%%%%%%%%%%%

Oscillons \cite{BM, G}, i.e., long-living, localized, quasiperiodic solutions of nonlinear field theories remain to be rather mysterious objects. Despite various applications \cite{G,CGM, Gleiser:2004an, G-cosm, Gr, Gr-2, Gleiser:2008ty, HS, GS-1, Am-1, FMPW, A, Am-2, Charukhchyan, Pujo, Dark-1, Lev, DS, LT, Aurrekoetxea:2023jwd, SYZ} and many years of investigations even the most basic features as their unexpectedly long lifetime and the characteristic modulations of the amplitude (double oscillations) are still a matter of a very active debate. 

Contrary to topological solitons or $Q$-balls  \cite{QB-1,QB-2,QB-3,QB-4}, the existence of oscillons is not directly related to any conserved (topological or nontopological) charge. Hence, their persistence to decay seems to rely on some hidden, not clearly identified properties of nonlinear field equations. In fact, it was advocated that there should exist a hidden relation to $Q$-balls with an approximately conserved $U(1)$ charge \cite{K, KT, I}. 

The well-known perturbative approach by Fodor, Forgacs, Horvath, and Lukacs (FFHL) \cite{Fodor:2008es} provides an approximated treatment of oscillons. However, this method is based on one parameter (frequency) and therefore fails to explain the modulation of the amplitude of the oscillations, which needs two independent degrees of freedom.  

Recently, it was proposed that an oscillon with modulations (excited oscillon) is a bound state of two unmodulated oscillons \cite{Blaschke:2024uec}. This was, however, based on a similarity of the analyzed model with the sine-Gordon theory, and in consequence, on an approximation of the modulated oscillon by a double breather.  

In the present work, we show that oscillons are indeed {\it closely related to $Q$-balls}, however, in a much more intriguing way than previously conjectured \cite{MT}. Namely, at the leading nonlinear order, all generic oscillons in (1+1) dimensions emerge from one {\it universal} $Q$-ball equation. This equation is obtained via Renormalization Group Perturbation Expansion (RGPE) \cite{Chen:1994zza, Chen:1995ena} and well coincides with the {\it integrable} complex sine-Gordon theory. This allows us to approximate {\it generic} modulated oscillons as a dressed bound state of two $Q$-balls, each defining its own unexcited oscillon. 
%%%%%%%%%%%%%%%%%%%%%%%%%%%%%%%%%%%%%
\section{RGPE method}
\label{sec:II}
%%%%%%%%%%%%%%%%%%%%%%%%%%%%%%%%%%%%%
The RGPE method introduced in  \cite{Chen:1994zza, Chen:1995ena} is in general executed as follows:
\begin{itemize}
\item Insert a naive perturbation series into the equations of motion that corresponds to small-amplitude expansions in fields. The leading order equations should be, therefore, linear wave equations corresponding to field fluctuations around a chosen vacuum.
\item Insert a monochromatic wave solution as a starting point of the series with some (typically complex) \emph{bare} amplitude $A_0$.
\item Solve the naive perturbation series order by order until a resonant (secular) term is encountered, indicating a breakdown of the perturbation expansion at a particular \emph{cutoff} scale.
\item Redefine the bare amplitude $A_0$ in terms of a \emph{dressed} amplitude $A$ in such a way that an artificial \emph{renormalization} scale is introduced, while the cutoff scale is absorbed into the definition of $A$.
\item Derive the renormalization group equations (RGEs) as consistency conditions that the solution is independent on the renormalization scale and on the particular form of the secular terms to a given order. 
\item Solve RGE and set the renormalization scale in such a way that all secular terms in the expansion are removed, i.e., choose a subtraction scheme. This gives a renormalized solution that is a global approximation of the true solution valid in some range of amplitudes. 
\end{itemize}

%%%%%%%%%%%%%%%%%%%%%%%%%%%%%%%%%%%%%
\section{Oscillon from $Q$-balls}
\label{sec:III}
%%%%%%%%%%%%%%%%%%%%%%%%%%%%%%%%%%%%%
Let us apply the RGPE algorithm to the relativistic (1+1)-dimensional scalar field theory with a generic potential having a minimum at $\phi=0$
\begin{equation}
V(\phi) = \frac{\phi^2}{2}-a_3\frac{\phi^3}{3}-a_4 \frac{\phi^4}{4}+...,
\end{equation}
where $a_3,a_4,... \in \mathbb{R}$. We insert a naive, small-amplitude perturbation expansion, i.e.
\begin{equation}
\phi = \varepsilon \phi_1 +\varepsilon^2 \phi_2 +\varepsilon^3 \phi_3 +\ldots  \,,
\end{equation}
into the equation of motion
$ \partial^2 \phi + \partial V / \partial \phi = 0$, 
where $\partial^2 \equiv \partial_t^2-\partial_x^2$. 
Here $\varepsilon$ is a book-keeping parameter.
This produces a hierarchy of equations 
\begin{subequations}
\begin{eqnarray}
\bigl(\partial^2 +1\bigr) \phi_1 & = & 0\,, \\
\bigl(\partial^2 +1\bigr) \phi_2 & = & a_3\phi_1^2\,, \\
\bigl(\partial^2 +1\bigr) \phi_3 & = & 2a_3\phi_1 \phi_2\, +a_4\phi_1^3, \\
\nonumber & \vdots &
\end{eqnarray}
\end{subequations}
Solving order by order we find
\begin{eqnarray}
\phi_1 &=&  A_0 \e^{\I \theta} + \cc \,, \\
\phi_2 &=& -\frac{a_3}{3}A_0^2 \e^{2\I \theta}+a_3A_0 \bar A_0 + \cc\,, \\
\phi_3 &=& \alpha A_0^3 \e^{3\I \theta}+\beta A_0^2 \bar A_0 \, \mathcal{S}(\theta,\bar\theta) \, \e^{\I \theta} + \cc\,,
\end{eqnarray}
where $\theta \equiv Q x-\Omega t$, $A_0$ is a bare amplitude and $\Omega = \sqrt{Q^2+1}$. Furthermore $\alpha= \frac{1}{24} \left(2a_3^2-3a_4 \right)$ and $\beta=10a_3^2/3 +3a_4$. We also introduced an auxiliary variable
$ \bar\theta \equiv \Omega x- Q t$. Finally, $ \mathcal{S}(\theta,\bar\theta)$ is the secular term, i.e. a solution to the equation
\begin{equation}
\label{eq:eqfors}
\bigl(\partial^2 +1\bigr)\Bigl( \mathcal{S}(\theta,\bar\theta) \e^{\I \theta}\Bigr) =  \e^{\I \theta}\,.
\end{equation}

Obviously, physics should not depend on the secular term  $\mathcal{S}$, as it is just an artifact of the perturbation expansion. Indeed, we will promote this statement as a guiding principle for removing the ambiguity in setting up RG equations.\footnote{The reader is invited to see \cite{BRSW} for further clarification on the secular term.}

Thus, let us continue to execute the RGPE algorithm. The bare solution to the third order in $\varepsilon$ reads
\begin{align}
\phi_B = &\ \varepsilon A_0 \e^{\I \theta} - \frac{\varepsilon^2a_3}{3}A_0^2 \e^{2\I \theta}+\varepsilon^2 a_3|A_0|^2 +\varepsilon^3 \alpha A_0^3 \e^{3\I\theta} \nonumber \\
&  + \varepsilon^3 \beta A_0 |A_0|^2\mathcal{S} \e^{\I\theta}+ \cc
\end{align}
Let us now define a dressed amplitude $A \equiv A(\theta_0, \bar\theta_0)$  in terms of renormalized scales $\theta_0$ and $\bar \theta_0$ as
\begin{equation}
A_0 = A\Big(1- \beta \varepsilon^2\mathcal{S}_0(\theta_0, \bar\theta_0)|A|^2+ \mathcal{O}(\varepsilon^3) \Big)\,.
\end{equation}
This leads to the bare solution
\begin{align}
\phi_B = &\ \varepsilon A \e^{\I \theta} - \frac{\varepsilon^2a_3}{3}A^2 \e^{2\I \theta}+\varepsilon^2 a_3 |A|^2 +\varepsilon^3 \alpha A^3 \e^{3\I\theta} \nonumber \\
&  +\varepsilon^3 \beta A |A|^2\bigl(\mathcal{S}-\mathcal{S}_0\bigr) \e^{\I\theta}+ \cc
\end{align}
The minimal subtraction scheme, where the entire secular term is removed once we set $\theta_0 = \theta$ and $\bar\theta_0 = \bar\theta$, is given by $\mathcal{S}_0 = \mathcal{S}$. 

\begin{figure}[thb]
\begin{center}
\includegraphics[width=0.99\columnwidth]{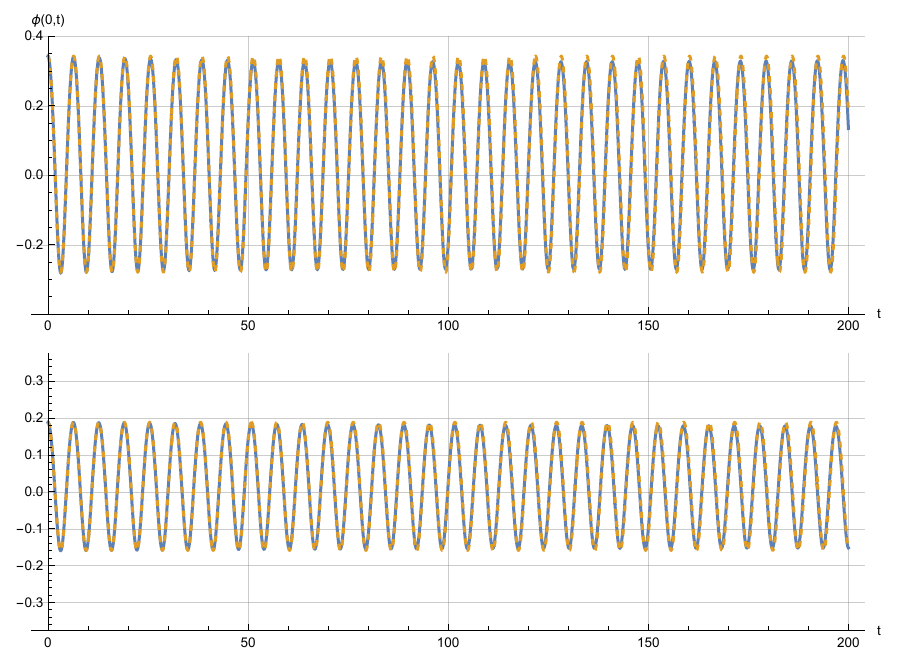}
\caption{\small Comparison between the unmodulated oscillon with (blue) a solution to full nonlinear equations of motion and renormalized solution (orange) for the single Q-ball solution given via Eq.~\refer{eq:qball} inserted into Eq.~\refer{eq:renormsol}, which is supplied as initial conditions. Upper: $\phi^3$ theory and $\lambda = 0.2$. Lower: double well $\phi^4$ theory and $\lambda = 0.15$. The oscillons are small and with negligible modulations. The FFHL approximation is indistinguishable from the orange curve. }
\label{fig:single}
\end{center}
\end{figure}

In order to remove the secular term, we need to ensure that the solution does not depend on the renormalization scales. 
However, a demand that
\begin{equation}
\frac{\partial \phi_B}{\partial \theta_0} = \frac{\partial \phi_B}{\partial \bar\theta_0} = 0 \,,
\hspace{4mm}
\forall\, x,\,t \,,
\end{equation}
leads to the RG equations
\begin{subequations}
\begin{eqnarray}
\frac{\partial A}{\partial \theta_0} &=& \beta \varepsilon^2 A \abs{A}^2 \frac{\partial\mathcal{S}_0}{\partial\theta_0}+ \mathcal{O}(\varepsilon^3) \,,
\\
\frac{\partial A}{\partial \bar\theta_0} &=& \beta \varepsilon^2 A \abs{A}^2 \frac{\partial\mathcal{S}_0}{\partial\bar\theta_0}+ \mathcal{O}(\varepsilon^3) \,,
\end{eqnarray}
\end{subequations}
that do depend on the $\mathcal{S}_0$.

Going with the minimal choice, $\mathcal{S}_0 = \mathcal{S}$, the only way to make the RG equation independent of the form of the secular term amounts to taking a combination of partial derivatives, i.e.,
\begin{gather}
2\I \frac{\partial A}{\partial \theta_0}+\frac{\partial^2 A}{\partial \theta_0^2}-\frac{\partial^2 A}{\partial \bar\theta_0^2} =\beta \varepsilon^2 A |A|^2 \Bigl(2\I \frac{\partial \mathcal{S}}{\partial \theta_0} 
+\frac{\partial^2 \mathcal{S}}{\partial \theta_0^2}  \nonumber \\  -\frac{\partial^2 \mathcal{S}}{\partial \bar\theta_0^2}\Bigr)
 =  \beta\varepsilon^2  A |A|^2\,, \hspace{4mm}
\forall\, x,\,t,\, \mathcal{S} \,,
\end{gather}
where in the last line we used Eq.~(\ref{eq:eqfors}). This can be further simplified by defining a new complex field
\begin{equation}
\Psi \equiv \sqrt{\frac{\beta}{2}}\varepsilon A \e^{\I \theta}\,,
\end{equation}
for which, at this order of perturbation, the RG equation takes the following {\it universal}  form
\begin{equation}\label{eq:RGrel}
\partial^2 \Psi +\Psi = 2 \Psi |\Psi|^2\,.
\end{equation}
This equation of motion follows from the complex $\phi^4$ Lagrangian
\begin{equation}\label{eq:lag1}
\mathcal{L} = |\partial \Psi |^2 -|\Psi |^2 +|\Psi |^4\,,
\end{equation}
and it possesses a stationary, $Q$-ball solution. To find it, we take $\Omega = 1$ and $Q=0$ so that $\theta = t$ and $\bar\theta =x$. Then, a single $Q$-ball solution of (\ref{eq:RGrel}) reads
\begin{equation}\label{eq:qball}
\Psi = \frac{\lambda }{\cosh(\lambda x)}\e^{\I \omega t}\,,
\end{equation}
where $\lambda$ is a scale parameter and $\omega = \sqrt{1-\lambda^2}$ is the frequency. 

The renormalized oscillon solution can be found via a {\it dressing} formula
\begin{equation}\label{eq:renormsol}
\phi_R = \sqrt{\frac{2}{\beta}}\Psi -\frac{2a_3}{3\beta}\Psi^2 +\frac{2a_3}{\beta} |\Psi|^2+\frac{\alpha2^{3/2}}{\beta^{3/2}}\Psi^3 +\cc \,,
\end{equation}
where the previously defined constants $\alpha$ and $\beta$ depend on the particular choice of the model.

This is our {\it first main result}. The oscillons emerge from $Q$-balls. Amazingly, {\it all generic oscillons}, (i.e., in models where $a_3$ or $a_4$ does not equal zero) independently of the details of the field theory, arise from the same $Q$-ball equations. Strictly speaking, this is true at the leading nonlinear order (generated at $\epsilon^3$ perturbation level). Higher-order corrections break this universality \cite{BRSW}. However, these are subleading corrections. Hence, oscillons in the $\phi^3$, the reverse $\phi^4$, the standard double well $\phi^4$, $\phi^6$, and many others \cite{Pujo-2} should be of the same universality class. Their properties, which is a well-known fact, are in many aspects very similar, see, e.g., \cite{Honda, Manton:2023mdr, We}.

Interestingly, there are theories with oscillons for which the RG equation does not have the upper form. This happens if $a_3=a_4=0$ and e.g., $a_6\neq 0$. Here we can include the exotic $\phi^6$ model with $V=\phi^2/2 - \phi^6/6$. The corresponding RG $Q$-ball equation has $\Psi |\Psi|^4$ leading nonlinearity. We conclude that oscillons in this model do not belong to the previous universality class and should have different properties. This is indeed the case \cite{BRSW}.

As an example, we consider $\phi^3$ model, where $a_3=1$ and all higher terms of the potential vanish  \cite{Manton:2023mdr} and the standard double well $\phi^4$ model with $a_3=3/2$ and $a_4=-1/2$. In Fig. \ref{fig:single} we compare the exact analytical RG approximation $\phi_R$ obtained from the {\it single} $Q$-ball solution with the actual oscillon generated from $\phi_R$ as an initial condition. For small amplitudes, the agreement is very good.  

\begin{figure}[thb]
\begin{center}
\includegraphics[width=0.99\columnwidth]{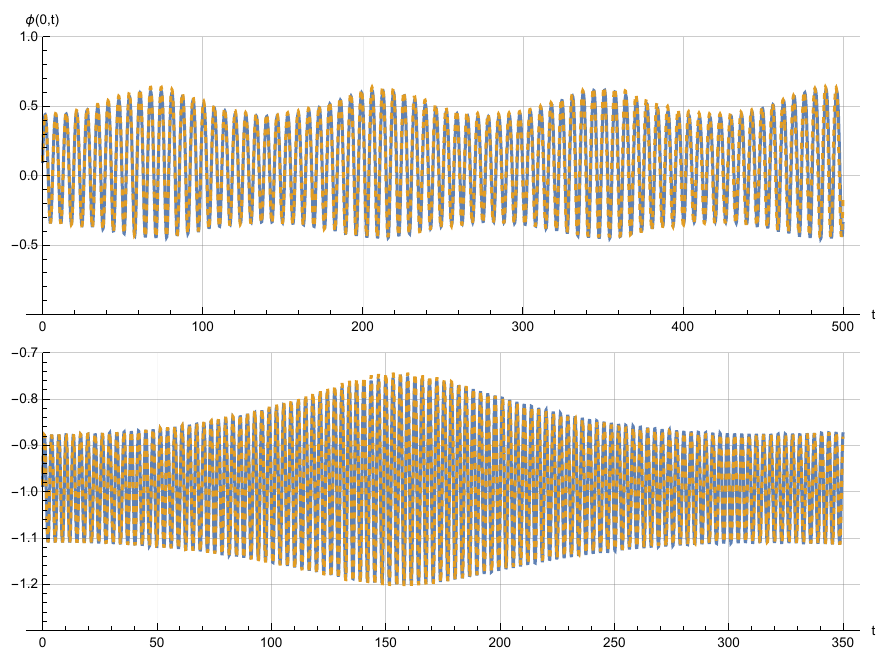}
\caption{\small Comparison between the modulated oscillon with (blue) a solution to full nonlinear equations of motion and renormalized solution (orange) for the two Q-ball solution given via Eq.~\refer{eq:twoQ} inserted into Eq.~\refer{eq:renormsol}, which is supplied as initial conditions. Upper: $\phi^3$ theory and $\lambda_1=0.3$, $\lambda_2=-0.05$. Lower: the double well $\phi^4$ theory and $\lambda_1=0.15$, $\lambda_2=-0.05$. The two Q-ball solution captures the characteristic amplitude modulation.}
\label{fig:two}
\end{center}
\end{figure}

We remark that our approximation differs slightly from the approximate solution generated by the FFHL expansion.  
However, if applied to the sine-Gordon model, our method order by order reproduces the breather solution. This requires higher order terms in the RGPE. The details are presented in \cite{BRSW}. 
%%%%%%%%%%%%%%%%%%%%%%%%%%%%%%%%%%%%%
\section{Integrability and Amplitude Modulations}
\label{sec:IV}
%%%%%%%%%%%%%%%%%%%%%%%%%%%%%%%%%%%%%
The next step is to observe that, around the vacuum, the universal RG $Q$-ball equation can be approximated by the {\it integrable complex sine-Gordon} ($\mathbb{C}$sG) equation
\begin{equation}
\bigl(\partial^2 +1\bigr)\Psi = \Psi |\Psi|^2 - \bar\Psi \frac{\partial_\mu \Psi \partial^\mu \Psi}{1- |\Psi|^2}\,.
\end{equation} 
The sense, in which this equation is ``close'' to RG equation \refer{eq:RGrel} is that it has exactly the same $Q$-ball solution given in Eq.~\refer{eq:qball}. Furthermore, the Lagrangian describing $\mathbb{C}$sG model differs from \refer{eq:lag1} by a multiplicative factor of the form $1+ \mathcal{O}\bigl(|\Psi|^2\bigr)$, i.e.,
\begin{equation}
\mathcal{L}_{\mbox{$\mathbb{C}$sG}} = \frac{1}{1-|\Psi|^2}\Bigl(\partial_\mu \bar\Psi \partial^\mu \Psi - |\Psi |^2+|\Psi |^4\Bigr)\,.
\end{equation}

\begin{figure*}
\begin{center}
\includegraphics[width=0.95\textwidth]{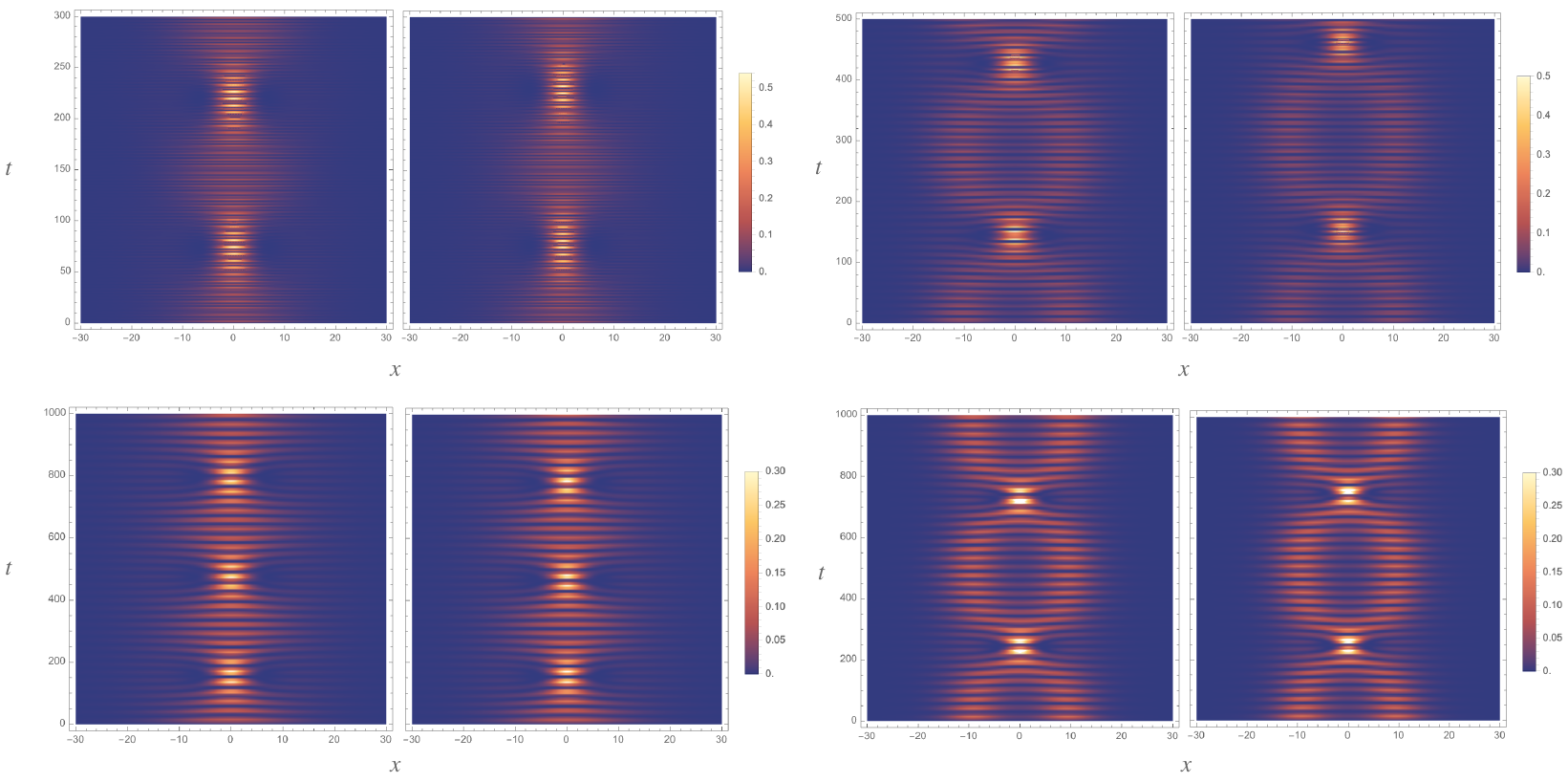}
\caption{\small Comparison for two Q-ball solution supplied as initial configuration. For better clarity, we plot $|\partial^2 \phi+\phi|$ as a function of $x$ and $t$ for both the numerical solution (left panel) and the renormalized solution (right panel). Upper: $\phi^3$ theory for $\lambda_1=0.3$, $\lambda_2=-0.1$ (left) and $\lambda_1=0.25$, $\lambda_2=-0.15$ (right). Lower: the double well $\phi^4$ theory for $\lambda_1=0.05$, $\lambda_2=-0.15$ (left) and $\lambda_1=0.1$, $\lambda_2=-0.15$ (right). As we see, the two Q-ball solution captures the characteristic amplitude modulation.}
\label{fig:two-den}
\end{center}
\end{figure*}

Importantly, due to the integrability the $\mathbb{C}$sG model supports a two Q-ball solution \cite{Bowcock:2008dn} describing a motion of two single $Q$-balls, each with its own scale parameter
\begin{widetext}
\begin{equation}\label{eq:twoQ}
\Psi_{12} = \frac{\I \bigl(\omega_1 -\omega_2\bigr)\Bigl(\frac{\lambda_1}{\cosh(\lambda_1 x)}\e^{\I \omega_1 t}-\frac{\lambda_2}{\cosh(\lambda_2 x)}\e^{\I \omega_2 t}\Bigr)}
{1-\omega_1\omega_2 -\lambda_1\lambda_2 \Bigl(\tanh(\lambda_1 x) \tanh(\lambda_2 x)+\frac{\cos(t(\omega_1-\omega_2))}{\cosh(\lambda_1 x)\cosh(\lambda_2 x)}\Bigr)}\,, \;\;\; \omega_{1,2}\equiv \sqrt{1-\lambda_{1,2}^2} \,.
\end{equation}
\end{widetext}
In Fig. \ref{fig:two} and Fig. \ref{fig:two-den}, we compare the approximated renormalized solution based on the two $Q$-ball and the actual modulated oscillon generated by $\phi_R$ as the initial configuration. We show results for the $\phi^3$ theory and the double well $\phi^4$ theory. The agreement is spectacular. 

This is our {\it second main result}. It is clearly visible that the {\it modulations are fully captured by the motion of two dressed $Q$-balls}. This supports the idea that a modulated oscillon is in fact a bound state of two unmodulated (fundamental) oscillons \cite{Blaschke:2024uec}. Here the actual bound state is generated from the two $Q$-ball solution. It also agrees with an observation that excited $Q$-ball coincides with two $Q$-solution \cite{Bowcock:2008dn}.

%%%%%%%%%%%%%%%%%%%%%%%%%%%%%%%%%%%%%
\section{Summary}
\label{sec:V}
%%%%%%%%%%%%%%%%%%%%%%%%%%%%%%%%%%%%%
In this work, we have resolved long-standing, fundamental problems related to oscillons.

First of all, we have established a nontrivial, RG-based, {\it relation between oscillons and $Q$-balls}. An intimate relation between these objects has been conjectured for a long time. We have shown that this is indeed the case. However, unexpectedly, the underlying (hidden) $Q$-balls, which are seeds for the oscillons, follow, at least at $\epsilon^3$ order, from a {\it universal complex field theory} derived via the RG flow. In other words, all generic oscillons emerge from the same $Q$-balls. Therefore, they should belong to the same universality class. We also identified other universality classes. They possess oscillons with qualitatively different properties. 

The simplest, single $Q$-ball provides an approximation very similar to the standard FFHL expansion \cite{Fodor:2008es} which very well reproduces dynamics of the simplest, unexcited i.e., unmodulated oscillons. However, due to a similarity of the RG oscillon equation to the integrable complex sine-Gordon theory, our approach naturally introduces a higher number of degrees of freedom associated with a higher number of $\mathbb{C}$sG  $Q$-balls. This is a great improvement in comparison with \cite{Fodor:2008es} allowing for an approximated analytical description of modulated oscillons. 

As a consequence, we have resolved the long-standing issue of the {\it origin of the amplitude modulation}, which is the most characteristic feature of oscillons. An excited, {\it modulated oscillon is a bound state of two unmodulated oscillons} originating in a two $Q$-ball solution. The modulation is an effect of a nonlinear superposition of these two oscillons each with its own DoF, that is frequency. This confirms the recent proposal presented in 
\cite{Blaschke:2024uec}. 

The observed relation between oscillons and $Q$-balls suggests that various $Q$-ball phenomena (like, e.g., charge swapping \cite{PS1} or superradiance \cite{PS2}) may possess oscillon counterparts.

Finally, the fact that the universal $Q$-ball equation in the generic universality class is well approximated by an {\it integrable} field theory opens a plausible and novel way to explain the longevity of oscillons. Here, the obvious reason is the integrability. Since the $Q$-ball equation is only approximated by an integrable theory oscillons eventually decay. Undoubtedly, this should be further studied. 

Approximated integrability gives also a path for investigation of quantum oscillons, which contrary to common belief \cite{H} may be long living objects \cite{J}. 

Although our work focused on (1+1) dimensional oscillons there are no obstacles to applying this framework for higher dimensions.

%%%%%%%%%%%%%%%%%%%%%%%%%%%%%%%%%%%%%
\acknowledgments
%%%%%%%%%%%%%%%%%%%%%%%%%%%%%%%%%%%%%
F.B. acknowledges the institutional support of the Research Centre for Theoretical Physics and Astrophysics, Institute of Physics, Silesian University in Opava and the support of Institute of Experimental and Applied Physics in Czech Technical University in Prague.
This work has been supported by the Grant No. SGS/24/2024 Astrophysical processes in strong gravitational and electromagnetic fields of compact object. K.S. acknowledges financial support from the Polish National Science Centre 
(Grant No. NCN 2021/43/D/ST2/01122).  A.W. acknowledges support from the Spanish Ministerio de Ciencia e Innovacion (MCIN) 
with funding from European Union NextGenerationEU (Grant No. PRTRC17.I1) and Consejeria de Educacion from JCyL through the QCAYLE project, as well as MCIN Project 1114 No. PID2020-113406GB-I0 and the Grant No. PID2023-148409NB-I00 MTM.

%\newpage 

\end{document}